\documentclass[lettersize,journal]{IEEEtran}

\usepackage{amsmath,amssymb,amsfonts}

\usepackage{newtxtext,newtxmath}

\usepackage[utf8]{inputenc} 
\usepackage[T1]{fontenc}    
\usepackage{textcomp}       

\usepackage{graphicx}
\usepackage[caption=false,font=normalsize,labelfont=sf,textfont=sf]{subfig}
\usepackage{booktabs}
\usepackage{threeparttable}
\usepackage{array}
\usepackage{colortbl}
\usepackage{multicol}
\usepackage{stfloats}

\usepackage[dvipsnames]{xcolor}

\usepackage{algorithm}
\usepackage{algorithmic}

\usepackage{verbatim}
\usepackage{url}
\usepackage{balance}

\usepackage{fancyhdr}
\def\BibTeX{{\rm B\kern-.05em{\sc i\kern-.025em b}\kern-.08em
    T\kern-.1667em\lower.7ex\hbox{E}\kern-.125emX}}
\usepackage{balance}
\begin{document}
\title{Continuous-variable Measurement Device Independent MIMO Quantum Key Distribution for THz Communications}
\author{Leixin Wu, Congtian Deng, Jiayu Pan, Lingtao Zhang, Yanyan Feng, Runbo Zhao, Yang Shen, \\Yuying Zhang, Jian Zhou
\thanks{The research was funded by the National Natural Science Foundation of China grant numbers	62201620, 62272483 and 62402435, the Science Fund for Distinguished Young Scholars of Hunan Province grant number 2023JJ10078, the Young Fund of the Natural Science Foundation of Hunan Province grant number 2023JJ41059, the Scientific research startup fund project of introduced talents of Central South University of Forestry and Technology grant number 2021YJ0050, the Ningbo Yongjiang Talent Programme grant number 2023A-398-G and Natural Science Foundation of Ningbo grant number 2024J205.\\
	\textit{(Corresponding authors: Lingtao Zhang)}
	\\Leixin Wu, Lingtao Zhang, Yanyan Feng and Yang Shen are with College of Electronic Information and Physics, Central South University of Forestry and Technology, Changsha {\rm 410004}, PR China (e-mail:874067703@qq.com; zhang@csuft.edu.cn; fengyanyanhenu@163.com; 2632804120@qq.com). \\
	Congtian Deng is with James Watt School of Engineering, University of Glasgow, Glasgow {\rm G12 8QQ}, UK (e-mail: timd56541@gmail.com).​​​​​​​​​​​​​​​​\\
	Jiayu Pan is with the School of Software Technology, Zhejiang University, Ningbo {\rm 315100}, PR China (e-mail: jiayupan26@zju.edu.cn)
	\\	Runbo Zhao, Yuying Zhang and Jian Zhou are with College of Computer and Mathematics, Central South University of Forestry and Technology,  Changsha {\rm 410004}, PR China (e-mail:1538038647@qq.com;3118454087@qq.com;13142153489@163.com).} }

\markboth{}%
{How to Use the IEEEtran \LaTeX \ Templates}

\maketitle

\begin{abstract}
Although multiple-input multiple-output (MIMO) terahertz (THz)  continuous-variable quantum key distribution (CVQKD) is theoretically secure, practical vulnerabilities may arise due to {detector} imperfections. This paper explores a CV measurement-device-independent (MDI) QKD system operating at THz frequencies within a MIMO framework. In this system, measurement is delegated to an untrusted third party, Charlie, rather than the receiver, eliminating all detector attacks and significantly enhancing the system’s practical security. Using {transmit-receive} beamforming techniques, the system transforms MIMO channels into multiple parallel lossy quantum channels, enabling robust key distribution between Alice and Bob. This study examines entanglement-based and prepare-and-measure protocols, deriving secret key rates for both asymptotic and finite code scenarios. Simulations reveal the critical role of multiple antenna configurations and efficient homodyne detection in mitigating free-space path loss and maximizing key rates. Results indicate that system performance is optimized at lower THz frequencies for long-range transmissions and higher frequencies for short-range applications. The proposed protocol offers a scalable solution for secure quantum communications in next-generation wireless networks, demonstrating potential for deployment in both indoor and outdoor environments.
\end{abstract}


\section{INTRODUCTION}
\IEEEPARstart{T}he fifth generation of mobile communication technology (5G) has now been extensively adopted in a variety of applications, prompting researchers to investigate new cases and solutions for beyond 5G (B5G) and 6G systems \cite{1,2,3,4,5}. The objective of these next-generation technologies is to achieve higher data transmission rates and  lower latency compared to 5G \cite{6,7,8}. The overarching objective is to facilitate the interconnectivity of multiple devices, enable real-time end-to-end communications and support advanced artificial intelligence applications \cite{3,4,5,6,7,8,9,10}.


 In this context, THz frequencies have emerged as a key enabler for short-range, high-capacity wireless links envisioned in 6G networks \cite{11,12}. While current systems predominantly operate in the microwave bands, the THz spectrum offers several compelling advantages—such as ultra-wide bandwidth, terabit-per-second data rates, and high spatial resolution—making it particularly well-suited for 6G communication scenarios, including ultra-fast wireless access, high-capacity backhaul, and secure device-to-device communication in dense urban or indoor environments \cite{13,14,15}. However, the deployment of THz communication systems faces critical challenges, notably high path loss and molecular absorption. These impairments significantly degrade signal quality over distance, making channel estimation and detection more susceptible to eavesdropping and tampering \cite{16,17}.

Quantum key distribution (QKD) provides a method for securely generating encryption keys between two users, with its security based on the principles of quantum physics \cite{18,19,20}. QKD has been demonstrated its ability to directly implement encryption for communication in 6G networks, including the  one-time-pad \cite{21,22}. QKD is categorized into discrete-variable (DV) QKD and continuous-variable (CV) QKD \cite{99,23,24}. DVQKD primarily encodes information through the polarization or phase of single-photon pulses and has found extensive application in areas such as national defense and satellite communications \cite{25,26}, while CVQKD systems are rapidly moving from the laboratory stage to field applications and prototype development \cite{27,28}. These systems are compatible with existing telecommunications technology and can employ multiplexing to enhance key rates, positioning them favorably for short-range wireless networks \cite{24,28}. 

However, many current QKD implementations rely on optical frequencies for point-to-point communications, for which precise positioning and tracking are required. This approach does not meet the mobility demands of B5G and 6G communication applications \cite{29,30}. A newly proposed THz CVQKD system for mobile devices addresses these challenges  \cite{31,32}. Unlike optical links, THz waves does not require meticulous alignment and can generate stable keys even in adverse weather conditions, such as fog and dust, ensuring reliable internet access in extreme environments  \cite{31}.

Recently, Kundu et al. have proposed a multiple-input multiple-output (MIMO) CVQKD scheme designed for the THz frequency band \cite{33}. This scheme employs transmit-receive beamforming techniques to transform the MIMO channel into parallel lossy quantum channels, improving both the secret key rate and the effective transmission distance compared to traditional single-antenna QKD schemes  \cite{33,34}. Furthermore, they have proposed a least-squares-based channel estimation method to improve the security and efficiency of these systems \cite{34}. The input-output relationship between Alice and Bob during the key generation phase  has been analyzed, while accounting for the additional noise introduced by channel estimation errors and detector noise  \cite{34}. Additionally,  to further improve the security of the MIMO QKD protocol, eavesdropping restrictions are also discussed \cite{35}. 

Building upon these advancements, a reconfigurable intelligent surface (RIS)-based wireless communication system has been proposed to further enhance the MIMO CVQKD system \cite{100}. In this setup, communication occurs via two paths: a direct path between Alice and Bob and a wireless path facilitated by the RIS \cite{100}. Simulations show that the RIS plays a crucial role in improving both the secret key rate and the transmission distance, particularly by ensuring secure communication  in scenarios where Eve attempts to measure the additional modes in the RIS-Bob channel. A comprehensive analysis of the associated channel estimation and secret key rate for the RIS is presented in \cite{101}, underscoring its potential to substantially enhance the performance of MIMO-based CVQKD systems.

In parallel, Liu et al. have recently investigated orthogonal frequency division multiplexing (OFDM) and orthogonal time frequency space (OTFS) based CVQKD systems operating over doubly selective THz fading channels, which are representative of high-mobility scenarios \cite{99999}. Their work introduces a multi-carrier framework integrated with low-density parity-check (LDPC) codes and a modified multi-dimensional reconciliation algorithm, demonstrating that OTFS-based MIMO CVQKD can effectively mitigate the Doppler-induced inter-carrier interference and outperform its OFDM counterpart in mobile wireless environments.  It highlight the importance of jointly considering advanced waveform design, powerful reconciliation algorithms, and MIMO techniques for enabling practical THz CVQKD under realistic wireless conditions.



 Despite the potential advantages of MIMO THz QKD, several practical challenges remain. First, atmospheric absorption in the THz frequency band, mainly caused by water vapor, imposes a major limitation on the system of achievable secret key rate per channel use \cite{33}. Second, imperfections in current detection devices render the system susceptible to detector-side eavesdropping attacks \cite{36}. To address these challenges, we propose a CV measurement-device-independent (MDI) MIMO THz QKD scheme. In this approach, an untrusted third party is introduced to perform the necessary measurements, allowing two users to establish secret keys without relying on their own detection devices. Consequently,  the MDI technique provides resilience against potential attacks targeting the detector-side channels. The effectiveness of MDI QKD has been demonstrated in conventional optical fiber networks \cite{36,37,1000} and its potential for strengthening 6G communications over atmospheric channels is particularly promising. Furthermore, the proposed scheme offers strong potential for enabling high-speed and secure wireless communication in both indoor and outdoor environments. The main contributions of this work are summarized as follows:
\underline{}
\begin{itemize}
\item[1)]We propose a  MDI MIMO CVQKD approach to improve the THz QKD system based on an transmit-receive beamforming scheme  using singular value
decomposition. A transmit-receive beamforming technique is employed by Alice, Bob and Charlie to transform the MIMO channels into multiplexed SISO channels.
\item[2)]We describe the communication procedures of the MDI MIMO QKD protocol, including the details of the prepare-and-measure (PM) and entanglement-based (EB) protocols. In addition, we analyze the influence of Gaussian collective attacks on the MIMO channel.
\item[3)]We derive the asymptotic and finite code key rates for the proposed CVMDI QKD protocol with reverse reconciliation (RR). The asymptotic analysis is performed under the assumption of infinite data block size, which provides an upper bound on the achievable secret key rate.   For finite code analysis, the maximum likelihood estimation (MLE) \cite{46,47,48} is employed instead of the least-squares method used in \cite{34}. In the MLE method, critical steps such as privacy amplification and channel estimation are considered, which are integral to the practical implementation of the MDI MIMO CVQKD protocol. The key parameters are estimated based on block size, resulting in more accurate parameter estimation. This approach has been widely adopted in different CVQKD systems \cite{105,103,104}.
\item[4)]Through extensive simulations, we evaluate the performance of the  MIMO QKD scheme against a baseline single-input single-output (SISO) scheme.  We also investigate the maximum transmission distance for the proposed protocol with different MIMO configurations in the range of 0.1-1 THz and explore the effect of the homodyne detection efficiency on the SISO and MIMO protocol. {Moreover, we compare the secret key rate and maximum transmission distance of the proposed protocol with the other MIMO QKD schemes and identify the frequency intervals over which a positive key rate can be maintained.}
\end{itemize}
{Notation}: $\mathbf{V^{\dagger}}$ is the conjugate transpose of $\mathbf{V}$, $\mathbf{0}_{A \times B}$ is an $A \times B$ all-zero matrix, N$(A,B)$ is a Gaussian distribution with mean $A$ and variance $B$, $\chi^2$ is a chi-square distribution, $\boldsymbol{I}=\begin{pmatrix}1&0\\0&1\end{pmatrix}$  and $\boldsymbol{Z}=\begin{pmatrix}1&0\\0&-1\end{pmatrix}$ are Pauli matrices, $N_{T_{A,B}}$ indicates that Alice and Bob have the same transmitter antenna node and $N_{R_{A,B}}$ indicates that Alice and Bob have the same receiver antenna node.
\section{SYSTEM MODEL}
\subsection{Channel model}
In MDI MIMO protocol, Alice and Bob communicate with Charlie through a MIMO THz wireless channel, where Alice and Bob are equipped with $N_{T_A}$ and $N_{T_B}$  transmitter antenna nodes and Charlie captures the THz waves transmitted by Alice and Bob with $N_{R_A}$ and $N_{R_B}$ receiver antenna nodes respectively. { In this paper, we assume that Charlie’s two antenna arrays are sufficiently separated in space to ensure negligible inter-array interference.}   The corresponding  MIMO THz channel matrix $\mathbf{H_A}\in\mathbb{C}^{N_{R_A}\times N_{T_A}}$ and $\mathbf{H_B}\in\mathbb{C}^{N_{R_B}\times N_{T_B}}$ can be obtained by \cite{33,38}
{
\begin{equation}
	\begin{aligned}
		&	\mathbf{H_A}=\sum_{l=1}^{L_A}\sqrt{\gamma_{A_l}}e^{j2\pi f_c\tau_l}\psi_{N_{R_A}}(\phi_{l}^{R_A})\psi_{N_{T_A}}^\dagger(\phi_l^{T_A}),\\&	\mathbf{H_B}=\sum_{l=1}^{L_B}\sqrt{\gamma_{B_l}}e^{j2\pi f_c\tau_l}\psi_{N_{R_B}}(\phi_{l}^{R_B})\psi_{N_{T_B}}^\dagger(\phi_l^{T_B}).
	\end{aligned}
\end{equation}}
where { $\psi_{N_{R_A}}$ and $\psi_{N_{R_B}}$, which express the array response vector of uniform linear array, are formulated as \cite{33}}
{
\begin{equation}
	\begin{aligned}
		&\psi_{N_{R_A}}\left(\theta\right)=\frac{1}{\sqrt{N_{R_A}}}\left[1,e^{j\frac{2\pi}{\lambda}d_{a}\sin\theta},\ldots,e^{j\frac{2\pi}{\lambda}d_{a}(N_{R_A}-1)\sin\theta}\right]^{T},\\&\psi_{N_{R_B}}\left(\theta\right)=\frac{1}{\sqrt{N_{R_B}}}\left[1,e^{j\frac{2\pi}{\lambda}d_{a}\sin\theta},\ldots,e^{j\frac{2\pi}{\lambda}d_{a}(N_{R_B}-1)\sin\theta}\right]^{T},
	\end{aligned}
\end{equation}}
 {where $d_a$ is inter-antenna spacing, $L_A$ and $L_B$ are multipath components,} $\gamma_{A_l}$ and $\gamma_{B_l}$ are the path losses of the $l$-th multipath between Alice, Bob and Charlie. When antenna element $G_a$ is set to 30 and Rayleigh roughness factor is set to 1, the path losses $\gamma_{A_l}$ and $\gamma_{B_l}$ for line-of-sight (LOS) and non-LOS (NLOS) component can be modeled by \cite{21}
\begin{equation}
	\begin{aligned}
		&		\gamma_{A_l}=900N_{R_A}N_{T_A}(\frac{\lambda}{4\pi d_{AC}})^210^{-\frac{\delta_1 d_{AC}}{10}},\\&	\gamma_{B_l}=900N_{R_B}N_{T_B}(\frac{\lambda}{4\pi d_{BC}})^210^{-\frac{\delta_2 d_{BC}}{10}}.
	\end{aligned}
\end{equation}
where $d_{AC}$ and $d_{BC}$ are the shortest transmission distance between the senders and receiver, {the $\delta_{1,2}$ are free-space path loss per kilometer, whose value is related to the transmission frequency $f_{c}$. The wavelength  signal  ${\lambda}$ is inversely proportional to the transmission frequency $f_{c}$.  Definitions of other parameters not directly relevant to this work can be found in \cite{34}.}
In the singular-value decomposition scheme, the channel matrixs are denoted as \cite{35}
\begin{equation}
	\begin{aligned}
		&\mathbf{H_A}=\mathbf{U_A\Sigma_A V_A^{\dagger}},\\&	\mathbf{H_B}=\mathbf{U_B\Sigma_B V_B^{\dagger}}.
	\end{aligned}
\end{equation}
where $\mathbf{U_A}\in\mathbb{C}^{N_{R_A}\times N_{R_A}}$, $\mathbf{U_B}\in\mathbb{C}^{N_{R_B}\times N_{R_B}}$, $\mathbf{V_A}\in\mathbb{C}^{N_{T_A}\times N_{T_A}}$ and $\mathbf{V_B}\in\mathbb{C}^{N_{T_B}\times N_{T_B}}$ are  unitary matrices and the associated matrixs $\Sigma_A$ and $\Sigma_B$ are formulated as \cite{37}
\begin{equation}
	\begin{aligned}
		& \Sigma_A=	\begin{bmatrix}
			\mathrm{diag}\left\{\sqrt{T_{A_1}},\ldots,\sqrt{T_{A_{r_1}}}\right\} & \mathbf{0}_{r_1\times(N_{T_A}-r_1)} \\
			\mathbf{0}_{(N_{R_A}-r_1)\times r_1} & \mathbf{0}_{(N_{R_A}-r_1)\times(N_{T_A}-r_1)}
		\end{bmatrix},\\& \Sigma_B=\begin{bmatrix}
		\mathrm{diag}\left\{\sqrt{T_{B_1}},\ldots,\sqrt{T_{B_{r_2}}}\right\} & \mathbf{0}_{r_2\times(N_{T_B}-r_2)} \\
		\mathbf{0}_{(N_{R_B}-r_2)\times r_2} & \mathbf{0}_{(N_{R_B}-r_2)\times(N_{T_B}-r_2)}
	\end{bmatrix}.
	\end{aligned}
\end{equation}
where  $r_1$ and $r_2$ are the rank of the MIMO channel matrices $\mathbf{H_A}$ and $\mathbf{H_B}$ ,  and $\sqrt{T_{A_x}}(x=1,2,\ldots,r_1)$ and $\sqrt{T_{B_y}}(y=1,2,\ldots,r_2)$ are the $i$-th nonzero singular value of the  matrix.

\begin{figure*}[!t]
	\centering
	\includegraphics[width=\linewidth]{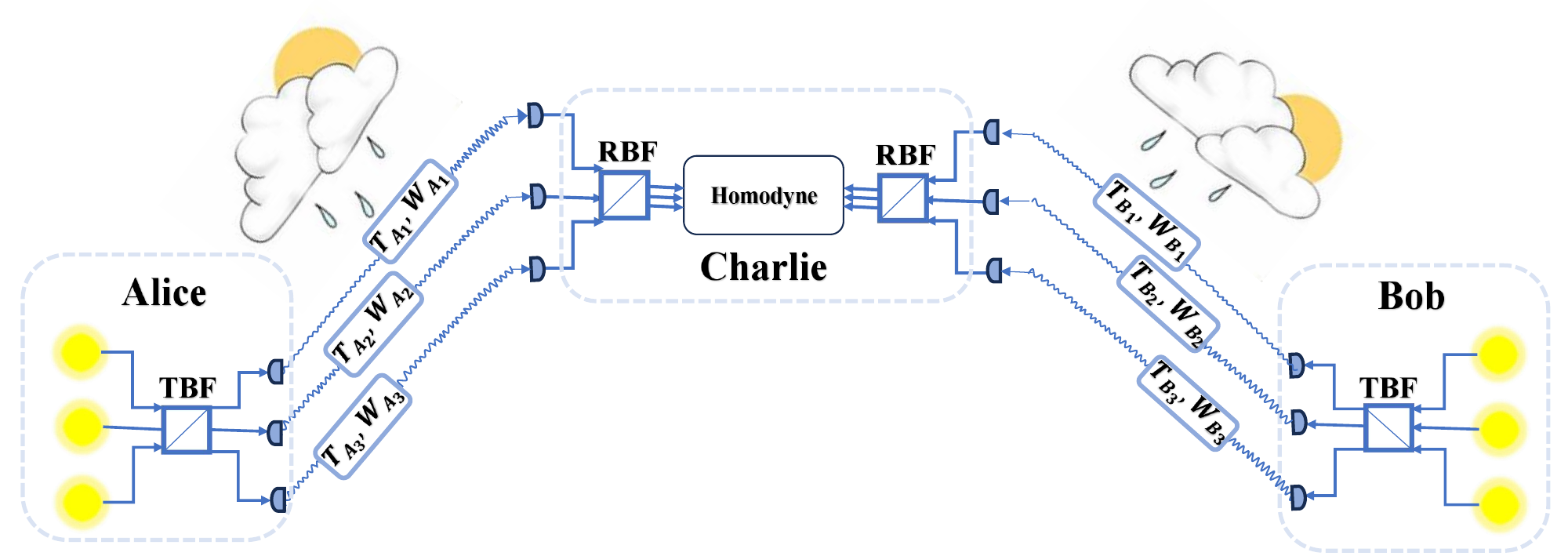}
	\caption{Schematic diagram of CVMDI protocol of the $3 \times 3$ MIMO configuration with the PM scheme  where Alice and Bob's transmitter antenna nodes are connected to Charlie's receiver antenna nodes via the free space channel with the transmission of the THz wireless channel ${T_{A_i}}$,${T_{B_i}}$ and  the noise variance  $W_{A_i}$,$W_{B_i}$.  {Alice (Bob) and Charlie employ transmit-receive beamforming techniques to decompose the original MIMO channel into multiple parallel SISO channels.} The TBF and RBF are transimit and receive beamforming, respectively.}
	\label{fig1}
\end{figure*}

\subsection{Prepare-and-measure scheme }
{The CVMDI MIMO protocol can be interpreted from two complementary perspectives. To explore the impact of MIMO antenna technology within the MDI framework, we first present the PM model, as illustrated in Figure \ref{fig1}. This model emphasizes the practical implementation of the protocol. In additon, we also present the EB equivalent model in the next section. These two representations are operationally equivalent and jointly provide a comprehensive understanding of the protocol. The corresponding steps of the PM scheme are as follows:}

\begin{itemize}
\item \emph{Step 1:}	Alice and Bob each select $r=$min$(r_1,r_2)$ random sequences $\{x_{A_i}$, $p_{A_i}\}$ $(i=1,2,\ldots,r)$ and $\{x_{B_i}$, $p_{B_i}\}$ from Gaussian distribution. They then prepare the corresponding $r$ coherent states $|x_{A_i}+ip_{A_i}\rangle$ and $|x_{B_i}+ip_{B_i}\rangle$, which are transmitted {using transmit beamforming techniques} through THz waves  from $N_{T_A}$ and $N_{T_B}$  transmitter antenna nodes to $N_{R_A}$ and $N_{R_B}$ receiver antenna nodes, respectively. The transmission of the THz wireless channel and the noise variance are  ${T_{A_i}}$,${T_{B_i}}$ and  $W_{A_i}$,$W_{B_i}$ respectively. 

\item \emph{Step 2:} {Charlie employs receive beamforming to collect the 2$r$ coherent states transmitted by Alice and Bob and subsequently performs Bell state measurement on the combined signals, producing 2$r$ output modes.} {The homodyne detection is used to measure the ${x}$-quadrature of Alice's mode and the ${p}$-quadrature of Bob's mode. The corresponding measurement outcomes, denoted by ${X_{C_i}}$ and $P_{{D_i}}$,  are subsequently forwarded to Alice's and Bob’s transmitter nodes for further processing \cite{37}}.
	\item \emph{Step 3:} After receiving the measurement results published by Charlie, Bob corrects his data: $X_{B_i}=g_iX_{C_i}+x_{B_i},P_{B_i}=g_iP_{D_i}-p_{B_i}$  {\cite{36}}, while Alice retains her original data $\{x_A,p_A\}$, where $g_i$ is the gain of the displacement operation. {Once Bob completes the data correction, entanglement swapping between Alice and Bob is effectively accomplished with the assistance of Charlie, resulting in strong correlations between their respective data.}
	\item \emph{Step 4:} Alice randomly selects portions of the retained data for parameter estimation and publishes them over a trusted channel. Bob uses this data to estimate parameters such as channel transmission, modulation variance and noise variance. He then evaluates the secret key rate. If the secret key rate is less than zero, they will terminate the session and initiate a restart of the key distribution.
\item \emph{Step 5:} Alice and Bob perform post-processing, including data reconciliation and privacy amplification, to obtain the final secret key rate. There are two methods for data reconciliation: direct reconciliation (DR) and  RR \cite{39}. Previous studies have shown that the DR scheme can only support short-distance QKD, so this paper focuses exclusively on the RR scheme \cite{40}.
\end{itemize}
{Alice and Bob employ orthogonal quadratures to achieve optimal entanglement swapping, thereby ensuring a strong correlation between Alice’s retained data and the corrected data obtained after Bob’s displacement operation \cite{36}. Moreover, the MDI MIMO protocol delegates the measurement to an untrusted third party instead of the receiver, effectively preventing detector-side attacks. As a result, the protocol remains secure with imperfect detectors.
}

\subsection{Gaussian collective attacks}
{In quantum communication, third-party eavesdroppers can attack quantum channels during transmission. Three main types of attacks are recognized: Gaussian individual attack, Gaussian collective attack and Gaussian coherent attack. In this paper, only the collective attack is discussed. According to quantum mechanics, Gaussian collective attacks are the most effective threat to CVQKD with standard RR protocols \cite{39}. Considering the CVQKD protocol under Gaussian collective attack can evaluate the security of the protocol.} To illustrate the effect of Gaussian collective attacks, the $2 \times 2$ MIMO channels are used as an example.  Based on Ref.  \cite{34}, the conclusion reached in the $2 \times 2$ model can be extended to the $N_{R_{A,B}} \times N_{T_{A,B}}$ MIMO model.  Alice and Bob’s two transmitted modes are first combined through a beam splitter. For the sake of simplicity, we assume that Alice and Bob transmit signals from the $j$-th transmitter node directly to Charlie's $j$-th receiver node. {The input-output relationship for the $j$-th beam splitter is defined as follows:} {\cite{35}}
{\begin{equation}
	\begin{bmatrix}\widehat{s}_{\mathrm{out},A(B)_{1}}\\\widehat{s}_{\mathrm{out},A(B)_{2}}\end{bmatrix}=\mathbf{B}_{T_{A(B)}}\begin{bmatrix}\widehat{s}_{\mathrm{in},A(B)_{1}}\\\widehat{s}_{\mathrm{in},A(B)_{2}}\end{bmatrix},
\end{equation}}
where the matrice of the beam-splitter are \cite{41}
\begin{equation}
	\mathbf{B}_{T_{A(B)}}=\begin{bmatrix}\sqrt{T_{A(B)}}&\sqrt{1-T_{A(B)}}\\-\sqrt{1-T_{A(B)}}&\sqrt{T_{A(B)}}\end{bmatrix}.
\end{equation}
Eve prepares $4$ auxiliary states consisting of EPR pairs with variance $W_{A(B)_j}$, and  $j=1,2$, denoted as $\{e_{A_j}, E_{A_j}\}$ and $\{e_{B_j}, E_{B_j}\}$, where the $E_{A_j}$ and $E_{B_j}$ are stored in the quantum memory, and the other modes $e_{A_j}$ and $e_{B_j}$ are mixed by the splitter and sent to Charlie.  Following the announcement of the relevant information by Alice and Bob  via a classical communication channel,  Eve retrieves the output modes  $\hat{e}_{A_j}$ and $\hat{e}_{B_j}$ and stores them in the quantum memory. Finally, Eve measures the auxiliary modes in the quantum memory, $\{\hat{e}_{A_j}, E_{A_j}\}$ and $\{\hat{e}_{B_j}, E_{B_j}\}$, to extract valuable information. This combined strategy enables the optimization of the amount of information acquired during the transmission process.
\section{SECURITY ANALYSIS }

\subsection{Entanglement-based scheme}
In the process of deriving the cryptographic key, an EB model is commonly used, which is equivalent to the PM model. Specifically:
\begin{itemize}
	\item \emph{Step 1:} Alice and Bob each prepare $r$ two-mode squeezed vacuum (TMSV) states. Each TMSV state is then split into two modes using a 50:50 beam splitter. {The modes $A_i$ and $B_i$ are retained locally by Alice and Bob, while the corresponding other modes $A_i^{'}$  and $B_i^{'}$  are transmitted to an untrusted third party, Charlie, via a MIMO channel} { employing transmit beamforming at the sender sides of Alice and Bob}. The variances of the two-mode squeezed states are defined as follows: for Alice's mode, $V_{A_i}=V_{AM_i}+V_{AO_i}$ $(i=1,2,\ldots,r)$;  for Bob's mode, $V_{B_i}=V_{BM_i}+V_{BO_i}$. The thermal noise variance is $V_{A(B)O_i}=2\bar{{n}}+1$ and $\bar{n}=\frac{1}{\exp(hf_c/k_BT_k)-1}$ \cite{42,43},  the $T_k$ is the atmospheric temperature and the $h$ and $k_B$ are the Planck and Boltzmann constants, respectively.
\item \emph{Step 2:} {Charlie employs receive beamforming to coherently combine the incoming signals  from multiple antennas. Subsequently, Bell-state measurements are performed on the received modes} { $A_i^{''}$ and $B_i^{''}$ from Alice and  Bob using homodyne detection,} { yielding measurement results  ${X_{C_i}}$ and  ${P_{D_i}}$.} The measurement results ${X_{C_i}}$ and  ${P_{D_i}}$  are then publicly reported.
\item \emph{Step 3:} When receiving the measurement results, { a displacement operation $D(\beta)$ is performed on Bob's retained mode to get mode  $B_i^{'''}$, which  becomes entangled with Alice’s mode  $A_i$ \cite{36},} where $\beta = {g_i}\sqrt{\frac{V_{B_i}+1}{V_{B_i}-1}}(X_{C_i} + iP_{D_i})$ {\cite{36}}. The homodyne detection is then employed to measure {mode $B_i^{'''}$}, yielding the measurement result $P_{B_i}$. Alice also measures her retained mode with homodyne detection, yielding the result $X_{A_i}$​.
\end{itemize}
The subsequent steps, including data correction, parameter estimation and post-processing, follow the same procedures as the PM model, resulting in the generation of a secure shared key.
\subsection{Asymptotic key analysis}
The asymptotic key rate of the proposed protocol is derived in this subsection. In this case, Alice and Bob can easily access their respective channel transmission and modulation variance without any reserved data for evaluation. We extend the method in \cite{33} to compute the key rate for the CVMDI MIMO protocol. The core idea of this approach is to transform the key of MIMO into the sum of the key rate of multiple SISO channels, which is given by \cite{34}
\begin{equation}
	K_{\mathrm{MIMO}}^{Ar}=\sum_{i=1}^rK_i^{A}, \qquad i = 1, 2,...,r,
\end{equation}
where $r$ is the minimum of the rank of the channel matrices $\mathbf{H_A}$ and $\mathbf{H_B}$. Since the proposed protocol employs a two-way scheme, the computation of its key rate can be quite challenging. Fortunately, previous work \cite{36} has identified that the covariance matrix used in a single-channel protocol can be effectively utilized to characterize the quantum state in the two-way protocol. This method enables us to determine the secret key limit of the proposed protocol under collective attacks.

When THz waves are transmitted in a MIMO channel, the equivalent one-way transmission $T_i$ of the CVMDI QKD protocol is \cite{37}
\begin{equation}
	T_i=\frac{1}{2}{g_i}^2T_{A_i}
\end{equation}
For the following derivation, it is necessary to characterize the relationship between Alice's and Bob's respective channel excess noise and noise variance \cite{42}
\begin{equation}
	\varepsilon_{A_i}=\frac{W_{A_i}(1-{T_{A_i}})-1}{T_{A_i}}+1
\end{equation}
and
\begin{equation}
	\varepsilon_{B_i}=\frac{W_{B_i}(1-{T_{B_i}})-1}{T_{B_i}}+1.
\end{equation}
Since Charlie’s detectors  are untrusted in the MDI protocol, the noise introduced by the two homodyne detections is attributed as excess noise. The total equivalent excess noise at the output is {\cite{37}}
\begin{equation}
	\begin{aligned}
	\varepsilon_{i}=&\frac{T_{B_{i}}}{T_{A_{i}}}(\sqrt{\frac{2}{T_{B_{i}}g^{2}}}\sqrt{V_{B_{i}}-1}-\sqrt{V_{B_{i}}+1})^{2}\\&+\varepsilon_{A_{i}}+\frac{2+(\varepsilon_{B_{i}}-2)T_{B_{i}}}{T_{A_{i}}}+\frac{1-\eta_{D_{{B}_{i}}}}{\eta_{D_{{B}_{i}}}}+\frac{1-\eta_{D_{{A}_{i}}}}{\eta_{D_{{A}_{i}}}}.
\end{aligned}
\end{equation}
To minimize excess noise, we set ${g_i}^2=[2(V_{{B_i}}-1)]/[T_{{B}}(V_{{B_i}}+1)]$. The ${T_i}$  and $	\varepsilon_i$ are rewritten as
\begin{equation}
\begin{aligned}	&T_{i}=\frac{T_{{A}_{i}}(V_{{B}_{i}}-1)}{T_{B_{i}}(V_{{B}_{i}}+1)},\\&\varepsilon_{i}=\varepsilon_{{A_i}}+\frac{2+(\varepsilon_{{B_i}}-2)T_{B_{i}}}{T_{{A}_{i}}}+\frac{1-\eta_{D_{{A}_{i}}}}{\eta_{D_{{A}_{i}}}}+\frac{1-\eta_{D_{{B}_{i}}}}{\eta_{D_{{B}_{i}}}}.\end{aligned}
\end{equation}
{The equivalent one-way noise variance can be given by \cite{42}
\begin{equation} 
	\quad\widehat{W}_i=\frac{T_i(\varepsilon_i-1)+1}{1-T_i}.
\end{equation}
After transmission and the impact of noise, the covariance matrix $\widehat{\gamma}_{AB_{i}^{'''}}$ is modified as \cite{44,1111}
\begin{equation}
	\begin{aligned}\widehat{\gamma}_{AB_{i}^{'''}}=\begin{pmatrix}V_{A_i}\boldsymbol{I}&\sqrt{T_i(V_{A_i}^2-1)}\boldsymbol{Z}\\\\\sqrt{T_i(V_\mathrm{A}^2-1)}\boldsymbol{Z}&[T_iV_{A_i}+(1-T_i)\widehat{W}_i]\boldsymbol{I}\end{pmatrix}\end{aligned}.
\end{equation}}
In the proposed protocol, we focus only on the RR scheme, as it demonstrates a higher secret key rate and longer transmission distance compared to the DR scheme. The secret key rate of the $i$-th parallel channel $K_i^{A}$ with RR is given by \cite{33}
\begin{equation}
	K_i^{A}(V_A,V_B,T_{A_i},T_{B_i},W_{A_i},W_{B_i})=\beta S(A_i:B_i)-I(B:E_i),
\end{equation}
where $\beta\in[0,1]$ is the efficiency of RR and $S(A_i{:}B_i)$ is the mutual information between the transmitted quantum states from Alice to Bob over the $i$-th channel, which is described by \cite{45}{
\begin{equation}
	S(A_i:B_i)=\frac{1}{2}\mathrm{log}_2\left[1+\frac{T_iV_{{AM}_i}}{\Lambda_i\left(V_{AO_i},\widehat{W}_i\right)}\right],
\end{equation}
where  $\Lambda_i(X,Y)=T_iX+(1-T_i)Y$.} The $I(B_i:E_i)$ denotes the Holevo bound, which quantifies the information accessible to eavesdropper Eve over the $i$-th channel and is described by {\cite{27}} {
\begin{equation}
	\begin{aligned}I\left(B_i {:} E_{i}\right)&=S(\rho_E)-\sum_{x_B}p(x_B)S(\rho_{E|x_B})\\&=f\left(\lambda_1^i\right)+f\left(\lambda_2^i\right)-f\left(\lambda_3^i\right)-f\left(\lambda_4^i\right),\end{aligned}
\end{equation}}
where 
\begin{equation}
	f(x)=\left(\frac{x+1}{2}\right)\log_2\left(\frac{x+1}{2}\right)-\left(\frac{x-1}{2}\right)\log_2\left(\frac{x-1}{2}\right).
\end{equation}
and $\lambda_{1-4}^i$ are the symplectic eigenvalues. When signal variance $V_{{A(B)M}_i}>>$ thermal noise variance $V_{A(B)O_i}$, $\lambda_{1-4}^i$ are quantified as { \cite{45} } 
{	\begin{equation}\label{eq}
		\begin{gathered}\lambda_{1}^{i}=\widehat{W}_i, \lambda_{2}^{i}=\Lambda_i(\widehat{W}_i,V_{A_i})\\\lambda_{3,4}^{i}=\sqrt{\frac{1}{2}\left(A\pm\sqrt{A^2-4B}\right)},\end{gathered}
	\end{equation}
 where
 \begin{equation}
 		\begin{gathered}A=\frac{V_{A_i}\widehat{W}_i\Lambda\left(\widehat{W}_i,V_{A_i}\right)+\widehat{W}_i\Lambda\left(\widehat{W}_iV_{A_i},1\right)}{\Lambda\left(V_{A_i},\widehat{W}_i\right)},\\	B=\frac{V_{A_i}\widehat{W}_i^{2}\Lambda\left(\widehat{W}_i,V_{A_i}\right)\Lambda\left(\widehat{W}_iV_{A_i},1\right)}{\Lambda^{2}\left(V_{A_i},\widehat{W}_i\right)},\end{gathered}\end{equation}}
Eventually, the expression for the secret key rate of the CVMDI protocol is calculated by
\begin{equation}
	\begin{aligned}
		K_{\mathrm{MIMO}}^{Ar}=&\sum_{i=1}^{r}\frac{1}{2}\mathrm{log}_{2}\left(1+\frac{T_{i}V_{AM}}{T_{i}V_{AO_i}+T_{i}(\varepsilon_{i}-1)+1}\right)\\&-f(\widehat{W}_i)-f(\Lambda_i(\widehat{W}_i,V_{A_i}))+f\left(\lambda_3^i\right)+f\left(\lambda_4^i\right).
	\end{aligned}
\end{equation}

\subsection{Finite code rate analysis}
The finite code phenomenon plays a critical role in QKD systems, mainly because it influences parameter estimation and privacy amplification. Moreover, in practical quantum communication, the block size is inherently finite, which means that the ideal performance predicted for infinite data block size cannot be achieved.  To make the simulation closer to reality, we extend the analysis of the finite code effect, originally studied in Gaussian and discrete modulation QKD protocols, to the CVMDI MIMO THz protocol \cite{46,102}. Taking into account the effect of finite code under collective attacks, the finite code rate expression for the CVMDI QKD protocol can be formulated as \cite{46,47}:
\begin{equation}\small
	K_{\mathrm{MIMO}}^{Fr}=\sum_{i=1}^{r}\frac{N}{M}[\beta K_{i}^{F}(V_{A},V_{B},T_{{LA}_i},T_{{LB}_i},W_{UA_i},W_{UB_i})-\Delta(N)]
\end{equation}
where $M$ is the total data block size exchanged by Alice and Bob in the $i$-th parallel channel and only $N$ data are valid encryption keys. The $\Delta(N)$ is linked to the privacy amplification process, whose value varies with the data block size \cite{104}.
\begin{equation}
	\Delta(N)=(2{dim}H_{X}+3)\sqrt{\frac{\log_{2}{(2/\tilde{\epsilon})}}{N}}+\frac{2}{N}\log_{2}(1/\epsilon_{PA}),
\end{equation}
where the $(2{dim}H_{X}+3)\sqrt{\frac{\log_{2}{(2/\tilde{\epsilon})}}{N}}$ represents the convergence speed of the independent and identically distributed minimum entropy and the $dimH_X=2$ is the dimension of the Hilbert space of the $x$ and $p$ vectors in the original key. The $\tilde{\epsilon}$ is the smoothness parameter and $\epsilon_{PA}$ is the failure probability of the privacy amplification. Their optimal values are both $10^{-10}$ \cite{46}. 

In this model, the block size parameter $l = M - N$ is estimated by sampling the relevant variables. We assume that before the detector, the signals sent by Alice and Bob in the $i$-th parallel channel are represented by $y_{A_i}^{\prime}$ and $y_{B_i}^{\prime}$ respectively. Since the variables of Alice, Bob and Charlie follow a Gaussian distribution, the following relationships exist between the data received by Charlie's node and the original data of Alice and Bob \cite{46,103}:

\begin{equation}
	\begin{aligned}
		&	y_{A_i}^{\prime}=t_{A_i}^{\prime}x_{A_i}+z_{A_i}, \\
		&y_{B_i}^{\prime}=t_{B_i}^{\prime}x_{B_i}+z_{B_i}.\\
	\end{aligned}
\end{equation}
where the $z_{A_i} $ and  $z_{B_i} $ follow a Gaussian distribution with mean 0 and unknown variance ${\sigma_{A_i}^{\prime}}^{2}=1+T_{A_i}\varepsilon_{A_i}$, ${\sigma_{B_i}^{\prime}}^{2}=1+T_{B_i}\varepsilon_{B_i}$ and  $t_{A_i}^{\prime}=\sqrt{T_{A_i}}$, $ t_{B_i}^{\prime}=\sqrt{T_{B_i}}$. Their MLE are respectively \cite{46}
\begin{equation}
	\begin{gathered}\hat{t}_{A_{i}}^{\prime}=\frac{\sum_{j=1}^{k}x_{A_{ij}}y_{A_{ij}}^{\prime}}{\sum_{j=1}^{k}x_{A_{ij}}^{2}},\hat{t}_{B_{i}}^{\prime}=\frac{\sum_{j=1}^{k}x_{B_{ij}}y_{B_{ij}}^{\prime}}{\sum_{j=1}^{k}x_{B_{ij}}^{2}},\\\hat{\sigma}_{A_{i}}^{\prime^2}=\frac{1}{k}\sum_{j=1}^{k}\left(y_{A_{ij}}^{\prime}-\hat{t}_{A_{ij}}^{\prime}x_{A_{ij}}\right)^{2},\\\hat{\sigma}_{B_{i}}^{\prime^2}=\frac{1}{k}\sum_{j=1}^{k}\left(y_{B_{ij}}^{\prime}-\hat{t}_{B_{ij}}^{\prime}x_{B_{ij}}\right)^{2},\end{gathered}
\end{equation}
where $\hat{t}_{A_i}^{\prime},\hat{t}_{B_i}^{\prime}\hat{\sigma}_{A_i}^{\prime^2}$ and $\hat{\sigma}_{B_i}^{\prime^2}$ of independent estimators follow the following distribution  \cite{46}

\begin{equation}
	\begin{gathered}
	\hat{t}_{A_{i}}^{\prime}\sim\mathrm{N}(t_{A_{i}}^{\prime},\frac{\sigma_{A_{i}}^{\prime2}}{\sum_{j=1}^{k}x_{A_{ij}}^{2}}),\hat{t}_{B_{i}}^{\prime}\sim\mathrm{N}(t_{B_{i}}^{\prime},\frac{\sigma_{B_{i}}^{\prime2}}{\sum_{j=1}^{k}x_{B_{ij}}^{2}}),\\\frac{l\hat{\sigma}_{A_i}^{\prime^2}}{\sigma_{A_i}^{\prime2}},\frac{l\hat{\sigma}_{B_i}^{\prime^2}}{\sigma_{B_i}^{\prime2}}\sim\chi^2(l-1).
\end{gathered}
\end{equation}

Due to the limit of block size $l$ and probability $\epsilon_{PE/2}$, Alice and Bob can evaluate these parameters at confidence intervals, which are given by {\cite{104}}
\begin{equation}
	\begin{gathered}t_{A_{i}}^{\prime}\in\left[\hat{t}_{A_{i}}^{\prime}-z_{\boldsymbol{\epsilon}_{PE}/2}\sqrt{\frac{\hat{\sigma}_{A_i}^{\prime^2}}{lV_{A}}},\hat{t}_{A_{i}}^{\prime}+z_{\boldsymbol{\epsilon}_{PE}/2}\sqrt{\frac{\hat{\sigma}_{A_i}^{\prime^2}}{lV_{A}}}\right],\\t_{B_{i}}^{\prime}\in\left[\hat{t}_{B_{i}}^{\prime}-z_{\epsilon_{PE}/2}\sqrt{\frac{\hat{\sigma}_{B_i}^{\prime^2}}{lV_{B}}},\hat{t}_{B_{i}}^{\prime}+z_{\epsilon_{PE}/2}\sqrt{\frac{\hat{\sigma}_{B_i}^{\prime^2}}{lV_{B}}}\right],\\\sigma_{A_{i}}^{\prime}{}^{2}\in\left[\hat{\sigma}_{A_i}^{\prime^2}-z_{\epsilon_{PE}/2}\frac{\hat{\sigma}_{A_i}^{\prime^2}\sqrt{2}}{\sqrt{l}},\hat{\sigma}_{A_i}^{\prime^2}+z_{\epsilon_{PE}/2}\frac{\hat{\sigma}_{A_i}^{\prime^2}\sqrt{2}}{\sqrt{l}}\right],\\\sigma_{B_i}^{\prime}{}^2\in\left[\hat{\sigma}_{B_i}^{\prime^2}-z_{\epsilon_{PE}/2}\frac{\hat{\sigma}_{B_i}^{\prime^2}\sqrt{2}}{\sqrt{l}},\hat{\sigma}_{B_i}^{\prime^2}+z_{\epsilon_{PE}/2}\frac{\hat{\sigma}_{B_i}^{\prime^2}\sqrt{2}}{\sqrt{l}}\right],\end{gathered}
\end{equation}
where $z_{\epsilon_{PE/2}}=6.5$ and $\epsilon_{PE}$ is the failure probability of the parameter estimation process. For any modulated variance $V_{A(B)}$, the first order derivative of $	K_i^{F}$ with respect to $t_{A(B)}$ and $\sigma_{A(B)_{i}}^2$ are related as follows:
\begin{equation}
	\frac{\partial 	K_i^{F}}{\partial t_{A(B)_i}}|_{V_{A(B)}}>0
\end{equation}
and
\begin{equation}
	\frac{\partial 	K_i^{F}}{\partial\sigma_{A(B)_{i}}^2}|_{V_{A(B)}}<0.
\end{equation}
To enhance the security of the protocol, we analyze one of the fastest scenarios, where Alice and Bob always estimate the lower bound of the $\hat{t}_{LA(B)_{i}}$ and the upper bound of  $\hat{\sigma}_{UA(B)_i}^{\prime^2}$. We have
\begin{equation}
	\begin{aligned}&\hat{t}_{LA_{i}}^{\prime}=\hat{t}_{A_{i}}^{\prime}-z_{\boldsymbol{\epsilon}_{PE}/2}\sqrt{\frac{\hat{\sigma}_{A_i}^{\prime^2}}{lV_{A}}},\\&\hat{t}_{LB_{i}}^{\prime}=\hat{t}_{B_{i}}^{\prime}-z_{\boldsymbol{\epsilon}_{PE}/2}\sqrt{\frac{\hat{\sigma}_{B_i}^{\prime^2}}{lV_{B}}},\end{aligned}
\end{equation}
and
\begin{equation}
	\begin{aligned}&\hat{\sigma}_{UA_i}^{\prime^2}=\hat{\sigma}_{A_i}^{\prime^2}+z_{\epsilon_{PE}/2}\frac{\hat{\sigma}_{A_i}^{\prime^2}\sqrt{2}}{\sqrt{l}},\\&\hat{\sigma}_{UB_i}^{\prime^2}=\hat{\sigma}_{B_i}^{\prime^2}+z_{\epsilon_{PE}/2}\frac{\hat{\sigma}_{B_i}^{\prime^2}\sqrt{2}}{\sqrt{l}}.\end{aligned}
\end{equation}
By association of the equations $\hat{\sigma}_{UA_i}^{\prime^2}=1+T_{A_i}\varepsilon_{UA_i}$,  $\hat{\sigma}_{UB_i}^{\prime^2}=1+T_{B_i}\varepsilon_{U_i}$, $\hat{t}_{LA_{i}}^{\prime}=\sqrt{T_{LA_{i}}}$ and $\hat{t}_{LB_{i}}^{\prime}=\sqrt{T_{LB_{i}}}$, we have
\begin{equation}
	\begin{gathered}
	T_{{LA}_i}=(\sqrt{T_{{A}_i}}-z_{\varepsilon_{PE}/2}\sqrt{\frac{1+T_{{A}_i}\varepsilon_{{A}_i}}{lV_A}})^2,\\T_{{LB}_i}=(\sqrt{T_{{B}_i}}-z_{\varepsilon_{PE}/2}\sqrt{\frac{1+T_{{B}_i}\varepsilon_{{B}_i}}{lV_B}})^2.
	\end{gathered}
\end{equation}
and
\begin{equation}
	\begin{gathered}
	\varepsilon_{UA_i}=\varepsilon_{A_i}+z_{\varepsilon_{PE}/2}\frac{(1+T_{A_i}\varepsilon_{A_i})\sqrt{2}}{T_{A_i}\sqrt{l}},\\\varepsilon_{UB_i}=\varepsilon_{B_i}+z_{\varepsilon_{PE}/2}\frac{(1+T_{B_i}\varepsilon_{B_i})\sqrt{2}}{T_{B_i}\sqrt{l}}.
	\end{gathered}
\end{equation}
The estimated noise variance $ W_{UA_i} $ and $ W_{UB_i} $  are given by \cite{42}
\begin{equation}
	\begin{gathered}
		W_{{UA}_i}=\frac{T_{{LA}_i}\varepsilon_{{UA}_i}}{1-T_{{LA}_i}}+1,\\	W_{{UB}_i}=\frac{T_{{LB}_i}\varepsilon_{{UB}_i}}{1-T_{{LB}_i}}+1.
	\end{gathered}
\end{equation}

When $T_{A(B)_i}$ are replaced by $T_{LA(B)_i}$ and  $W_{A(B)_i}$ by $W_{UA(B)_i}$, using an similar approach with asymptotic key rate, we derive the finite code key rate $K_{\mathrm{MIMO}}^{Fr}$ for the CVMDI MIMO protocol.

\section{SIMULATION AND PERFORMANCE EVALUTION}

Next, an evaluation of the performance of both MIMO and SISO CVMDI protocols over atmospheric channels is conducted. To achieve the maximum spatial multiplexing gain, we set  multipath components $L_A=\min(N_{T_A},N_{R_A})$ and $L_B=\min(N_{T_B},N_{R_B})$ \cite{1010,1011}. In this case, The channel matrices $\mathbf{H_A}$ and $\mathbf{H_B}$ are full-rank matrices. The analysis is centered on a symmetric scenario, where the number of transmitter antenna nodes $N_{T_{A,B}}$ is equal to the number of receiver antenna nodes $N_{R_{A,B}}$, the MIMO channel connecting Alice and Charlie is identical to the one between Bob and Charlie and Charlie is in the middle of Alice and Bob. The total transmission distance is $d_{AB}=d_{AC}+d_{BC}=2d_{AC}$. {There exists a correlation between atmospheric loss and THz frequency. Accordingly, in this paper, we assume atmospheric losses $\delta_{1,2}$ are assumed to be 0.6 dB/km, 5 dB/km, 50 dB/km and 100 dB/km for the frequency bands 0.1 THz, 0.25 THz, 0.5 THz and 1 THz, respectively \cite{24,45}. Detailed values and definitions of the relevant parameters are provided in Table I.} 

\begin{figure}
	\centering
	\includegraphics[width=0.8\linewidth]{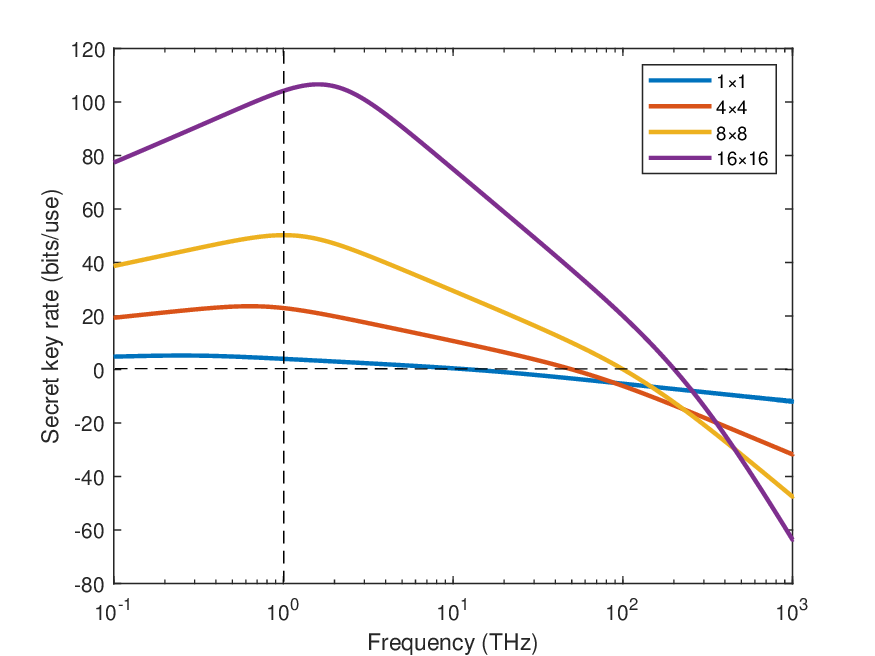}
	\caption{{Key rate as a function of operating frequency for multiple MIMO configurations at room temperature when transmission distance approaches zero.}}
	\label{fig10}
\end{figure}

{We first identify the conditions under which the proposed protocol achieves a positive secret key rate. Figure \ref{fig10} shows the secret key rate versus frequency for different MIMO configurations at room temperature when the transmission distance approaches zero. As the frequency increases moderately from 0.1 THz to 1 THz, the initial key rate(bits/channel use)  of the CVMDI MIMO QKD protocol improves. However, further increases in frequency lead to a significant degradation in system transmission, resulting in a reduction in the key rate. Notably, for frequencies above 300 THz, the protocol fails to generate a positive key rate. The subsequent analysis focuses on the frequency range of [0.1 THz, 1 THz]. Both the simulation of the asymptotic key rate and finite code key rate are considered.}
\begin{table*}
	\caption{\textbf{Relevant parameters}}
	\centering
	\begin{tabular}{cccc}
		\toprule
		Order&Symbol&Description&Value \\
		\midrule
		1&$N_{T_{A,B}}$&Transmitter antenna node&Independent variable \\
		2&$N_{R_{A,B}}$& Receiver antenna node&Independent variable \\
		3&$L_{A(B)}$&multipath components&Independent variable \\
		4&$V_{A(B)M}$&Modulation variance&$100000$ \cite{37} \\
		5&$\beta$&Reconciliation efficiency&1 \\
		6&$W_{A(B)_i}$&Noise variance&1 \cite{41} \\
		7&$\delta_{1,2}$&Atmospheric loss&Independent variable \\
		8&$k_B$& Boltzmann constant& $1.38 \times 10^{-23}$ \\
		9&$h$& Planck constant&$ 6.626\times 10^{-34}$ \\
		10&$f_c$& Frequency& Independent variable \\
		11&$T_k$&Temperature&$300$K \\
		12&$\eta_{D_{A(B)_{i}}}$&Detector efficiency&Independent variable \\
		13&$G_a$&antenna element&30 \cite{21}\\
		14&$M$&Block size&Independent variable \\
		15&$N$&Valid encryption keys&Independent variable \\
		16&$dimHx$&Hilbert space dimension&2 \cite{46}\\
		17&$\tilde{\epsilon}$&Smoothness parameter&$10^{-10}$ \cite{46}\\
		18&$\epsilon_{PA}$&Failure probability of privacy amplification&$10^{-10}$ \cite{47}\\
		19&$z_{\epsilon_{PE/2}}$&Probability error&$6.5$ \cite{46}\\
		20&$l$&Block size for parameter estimation&Independent variable \\
		\bottomrule
	\end{tabular}
\end{table*}

\subsection{Asymptotic key rate simulation}

\begin{figure*}[ht]
	\subfloat[ $f_c=100\mathrm{GHz}$]{\includegraphics[width=0.24\textwidth]{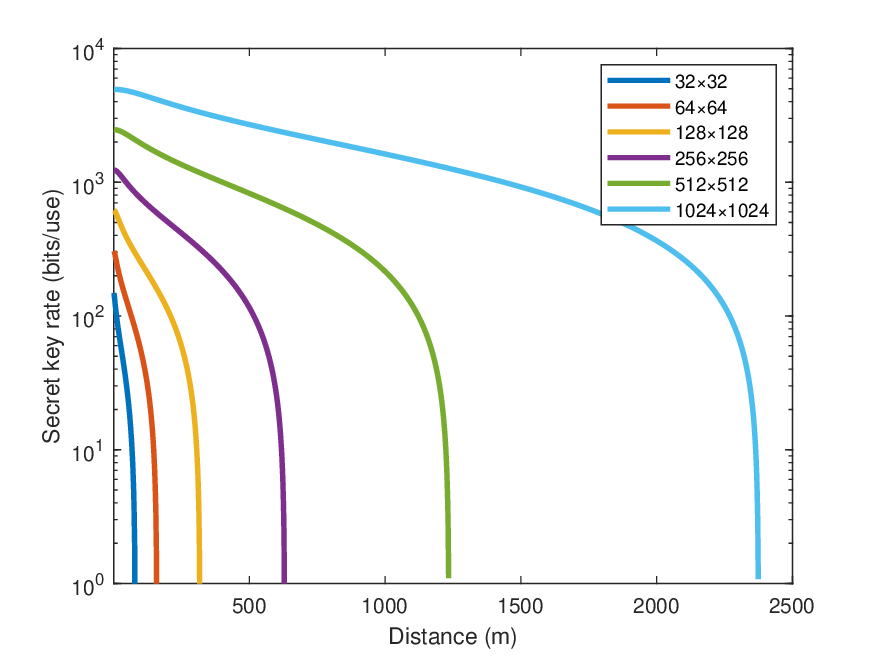}}
	\hfill 	
	\subfloat[$f_c=250\mathrm{GHz}$]{\includegraphics[width=0.24\textwidth]{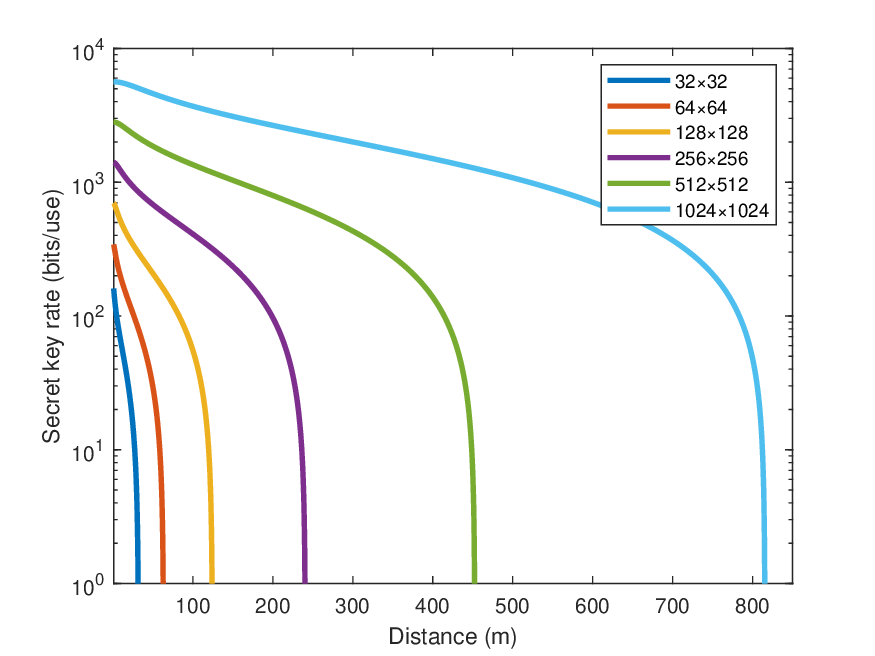}}
	\hfill 	
	\subfloat[$f_c=500\mathrm{GHz}$]{\includegraphics[width=0.24\textwidth]{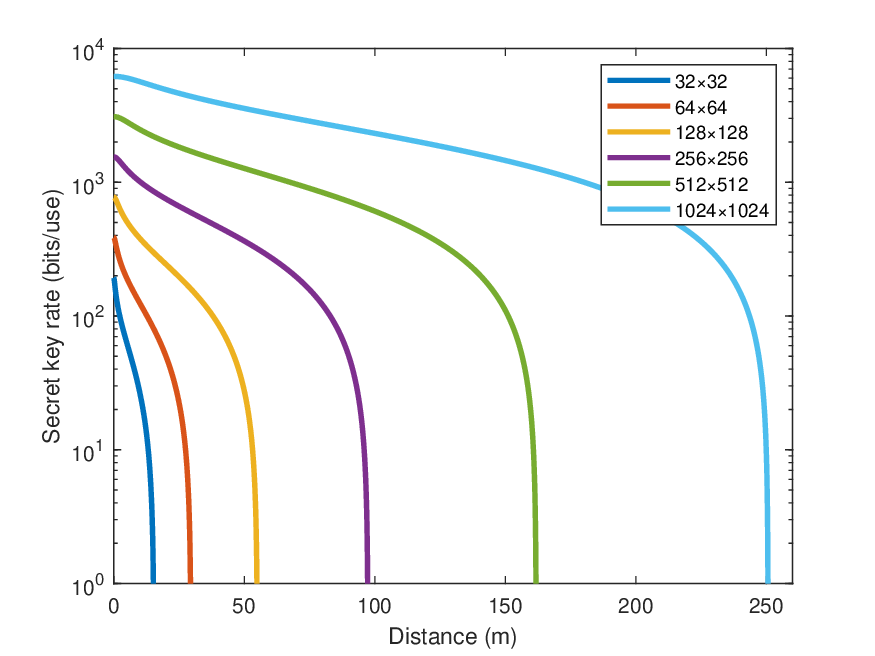}}
	\hfill	
	\subfloat[$f_c=1000\mathrm{GHz}$]{\includegraphics[width=0.24\textwidth]{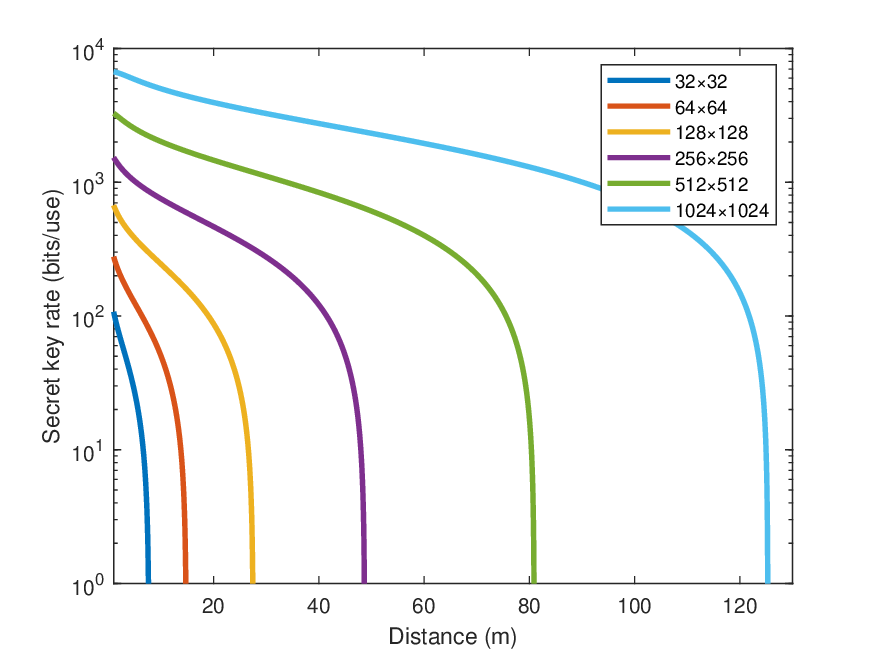}}
	\hfill 	
	\caption{Asymptotic key rate as a function of the transmission distance for the various MIMO configurations with different frequencies. As the frequency decreases from 1THz to 0.1THz, the maximum transmission distance in $1024\times 1024$ MIMO configuration gradually increases from  {125m to 2374m}. This trend also applies to other MIMO configurations.   $\eta_{D_{A(B)_{i}}}=1$, {atmospheric loss $\delta_{1,2}$: 0.6 dB/km for 0.1THz, 5 dB/km for 0.25THz, 50 dB/km for 0.5THz and 100 dB/km for 1THz}.}
	\label{fig2}
\end{figure*}
\begin{figure*}[ht]
	\subfloat[ $f_c=100\mathrm{GHz}$]{\includegraphics[width=0.3\textwidth]{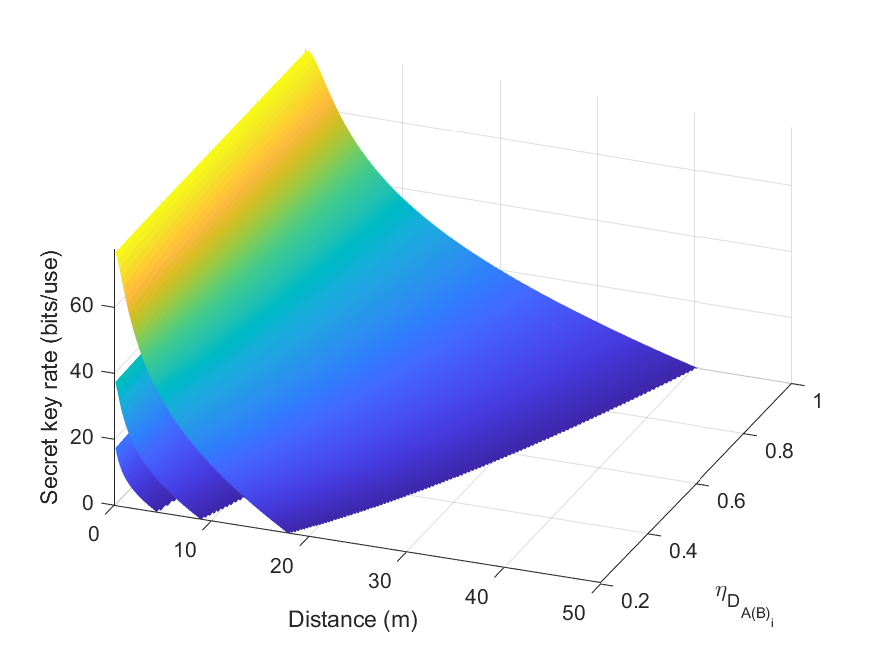}}
	\hfill 	
	\subfloat[$f_c=250\mathrm{GHz}$]{\includegraphics[width=0.3\textwidth]{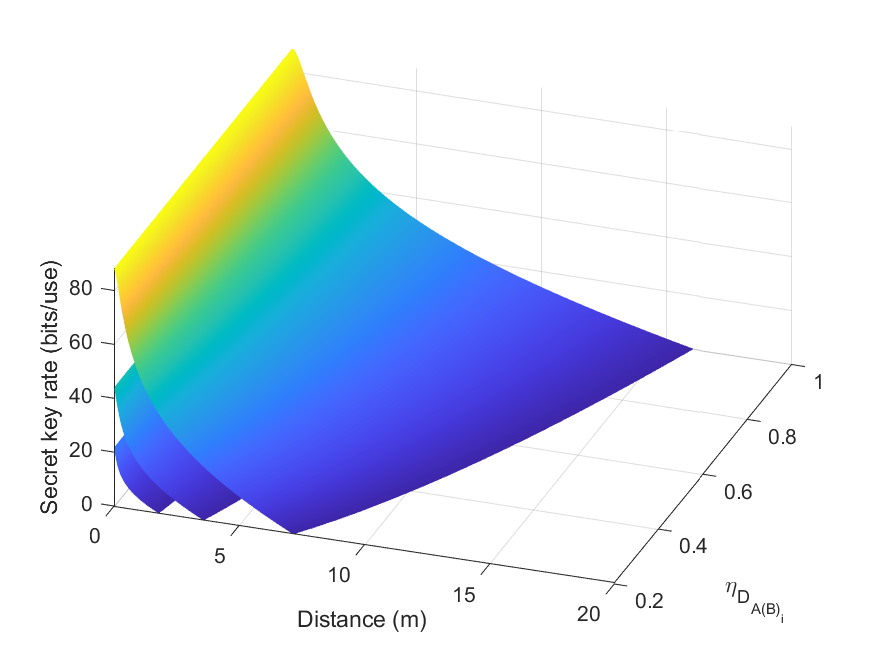}}
	\hfill 	 
	\subfloat[$f_c=500\mathrm{GHz}$]{\includegraphics[width=0.3\textwidth]{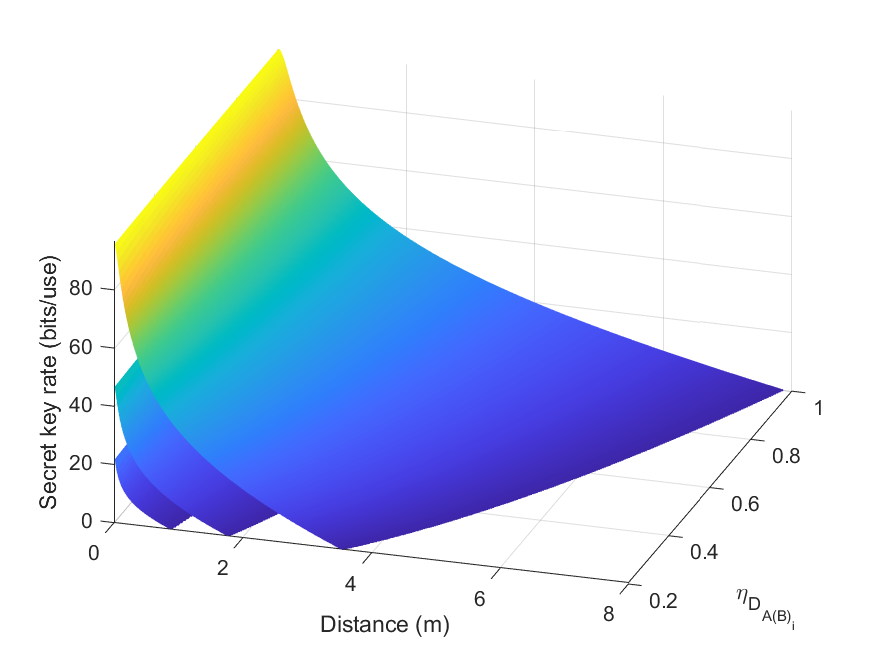}}
	\hfill	
	\caption{A three-dimensional surface plot showing the relationship between the $	K_{\mathrm{MIMO}}^{Ar}$, homodyne detector efficiency and the transmission distance with different frequencies. All subplots, arranged from top to bottom, represent the  $16\times16$, $8\times8$ and $4\times4$ MIMO configurations, respectively. {atmospheric loss $\delta_{1,2}$: 0.6 dB/km for 0.1THz, 5 dB/km for 0.25THz and 50 dB/km for 0.5THz}}
	\label{fig3}
\end{figure*}
\begin{figure}[!t]
	\centering
	\includegraphics[width=0.8\linewidth]{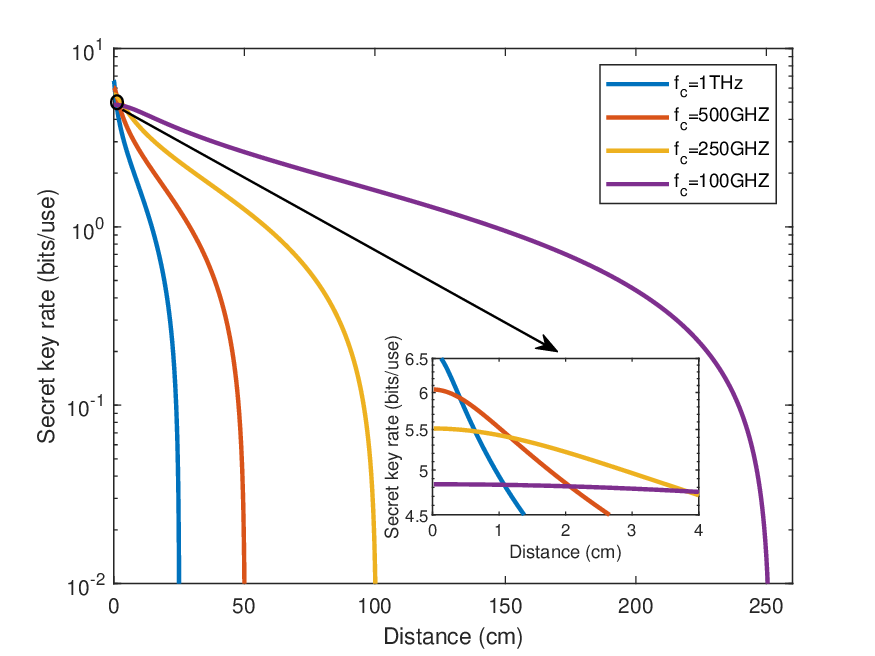}
	\caption{Secret key rate as a function of transmission distance for different frequencies in SISO protocol. It can be observed that, at shorter distances, higher frequency THz waves yield higher key rates. $\eta_{D_{A(B)_{i}}}=1$, {atmospheric loss $\delta_{1,2}$: 0.6 dB/km for 0.1THz, 5 dB/km for 0.25THz, 50 dB/km for 0.5THz and 100 dB/km for 1THz}.}
	\label{fig4}
\end{figure}
\begin{figure}[!t]
	\centering
	\includegraphics[width=0.8\linewidth]{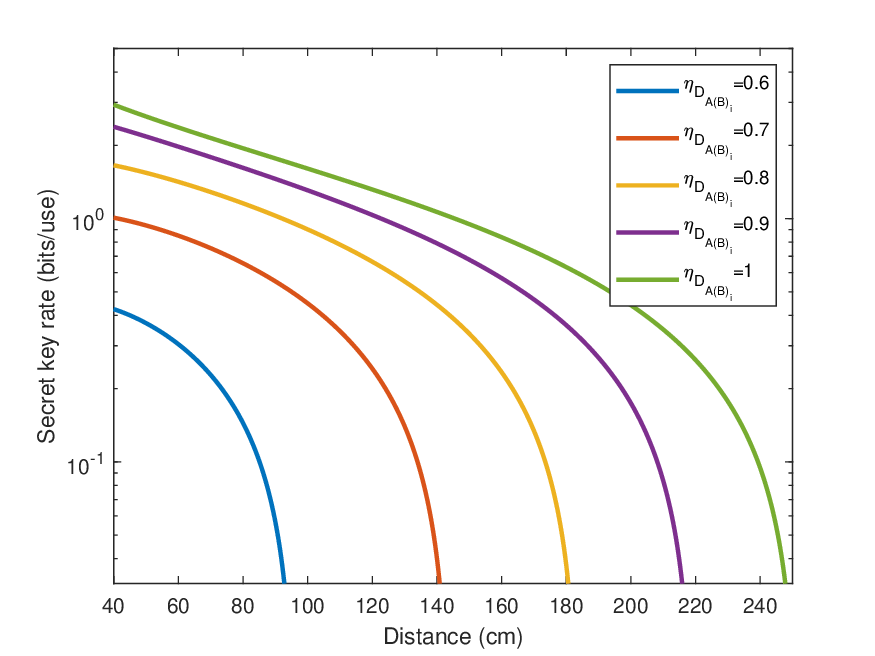}
	\caption{Asymptotic key rate as a function of transmission distance for different homodyne detector efficiency in the SISO protocol. The maximum transmission distance is less than half of the optimal case when the detector efficiency is 0.6. $f_c=100$GHz,  {$\delta_{1,2}=0.6$dB/km}.}
	\label{fig5}
\end{figure}


{
In the asymptotic regime, we consider an ideal scenario in which key distribution is error-free, and both communicating parties can achieve their respective channel transmissions and modulation variances without requiring additional data for parameter estimation. This assumption provides an upper bound on the secret key rate for the proposed protocol. Based on the derived expression in Eq. (22), we can evaluate the performance of the MDI MIMO THz QKD protocol. The relevant simulation parameters of the asymptotic key are presented in rows 1 through 13 of Table I.}

In Figure \ref{fig2}, we investigate the maximum transmission distance as a function of the asymptotic key rate 	$K_{\mathrm{MIMO}}^{Ar}$  over frequencies ranging from 0.1 THz to 1 THz for various MIMO configurations $(N_{T_{A,B}} \times N_{R_{A,B}} )$. Our results show that MIMO techniques significantly improve both the key rate and the maximum achievable transmission distance.  {According to Figure \ref{fig2} (a), (b), (c), (d), we can find the CVMDI QKD system performs optimally at 100 GHz for long distance transmission,} where the interplay between thermal noise, transmission and the frequency of THz wave is most favorable. {When the frequency decreases from 1 THz to 0.1 THz, thermal noise increases, which negatively affects the secret key rate. However, the transmission efficiency improves and the path loss is reduced, both of which are beneficial for key generation. The positive impact outweighs the negative impact of increased noise, resulting in an overall enhancement of the transmission distance. Consequently, we conclude that the optimal frequency for achieving the maximum transmission distance of CVMDI MIMO protocol lies at the lower limit of the THz channel  (0.1 THz). By appropriately adjusting the number of antennas and the frequency of the THz wave, proposed protocol can meet the requirements for both short-distance indoor and long-distance outdoor wireless communication.}

{In Figure \ref{fig3}, a three-dimensional surface plot shows the relationship between the key rate, homodyne detection efficiency $\eta_{D_{A(B)_{i}}}$ and maximum transmission distance for different MIMO configurations. In each subplot, the MIMO configurations are represented from top to bottom as $16\times 16$, $8\times8$ and $4\times4$, respectively. The results indicate that insufficient detector efficiency significantly degrades both the asymptotic key rate and the achievable transmission distance. This highlights the critical role of near-ideal detectors in enabling high-rate, long-distance key distribution, as poor detection performance can severely compromise key stability and overall protocol reliability—an issue largely overlooked in prior MIMO QKD studies.}

When $N_{T_{A,B}} \times N_{R_{A,B}}$ configuration is set to $1\times1$, the MIMO protocol is changed to the SISO protocol. Figures \ref{fig4} and \ref{fig5} provide a quantitative evaluation of the SISO protocol. Figure \ref{fig4} shows how the key rate varies with the transmission distance in different THz scenarios.  In comparison to larger MIMO configurations, the SISO protocol demonstrates suboptimal performance, with a maximum transmission distance of { almost} 250 cm. However, an interesting observation from Figure \ref{fig4} is that the SISO protocol exhibits a positive correlation between key rate and frequency at short ranges (0-4cm). {This finding serves as a critical note: it shows that higher frequencies play a key role in short-distance transmission, potentially improving the key rate for indoor communications. Conversely, for long-distance outdoor communications, it is imperative to maintain the frequency of the THz wave at low frequency (100 GHz) to ensure optimal performance.}

Figure \ref{fig5} further investigates the effect of homodyne detector efficiency on the SISO protocol at 100 GHz. In Figure \ref{fig5}, the conclusion can be observed that as the detector efficiency decreases from its optimal value to 0.6, the effective maximum transmission distance also decreases from {248 cm to 92 cm}. The decrease in transmission distance is primarily due to the increase in excess noise. Meanwhile, in Figure 4(a), we find that the rate of decrease of the MIMO protocol with a 16×16 configuration is significantly lower than that of the SISO protocol within the same interval [0.6, 1]. It indicates that the MIMO technique improves the resistance of the MDI protocol to noise variance, underscoring the merits of the MIMO approach.
\subsection{Finite code rate simulation}
\begin{figure*}[t]
	\subfloat[ $N_{T_{A,B}} \times N_{R_{A,B}} =8\times8$]{\includegraphics[width=0.24\textwidth]{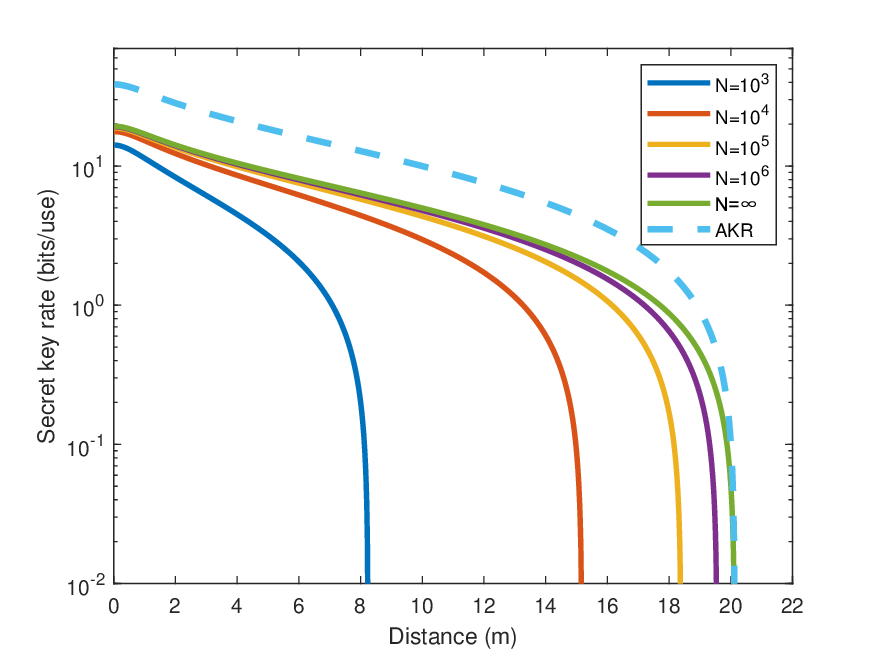}}
	\hfill 	
	\subfloat[$N_{T_{A,B}} \times N_{R_{A,B}}=16\times16$]{\includegraphics[width=0.24\textwidth]{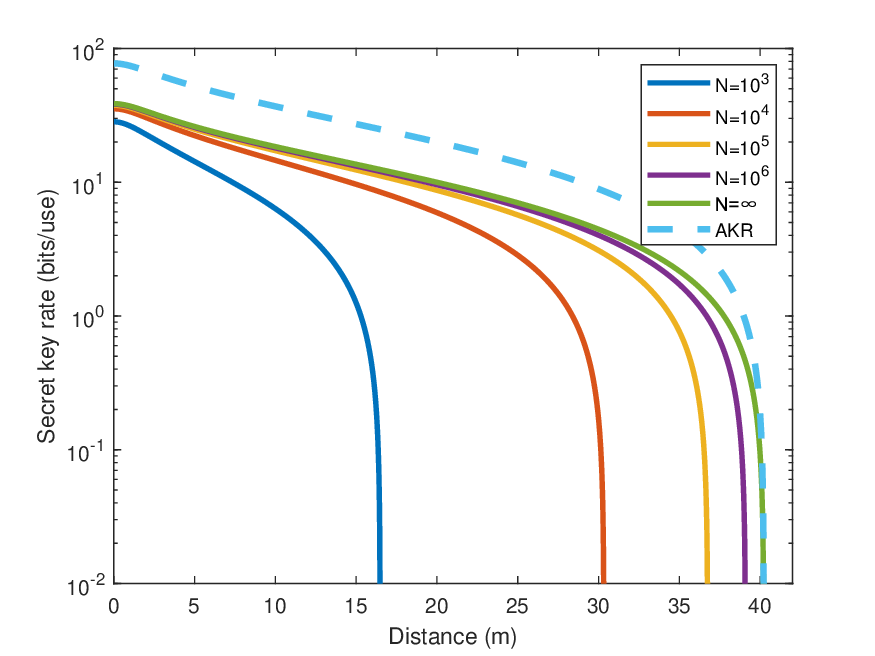}}
	\hfill 	
	\subfloat[$N_{T_{A,B}} \times N_{R_{A,B}}=32\times32$]{\includegraphics[width=0.24\textwidth]{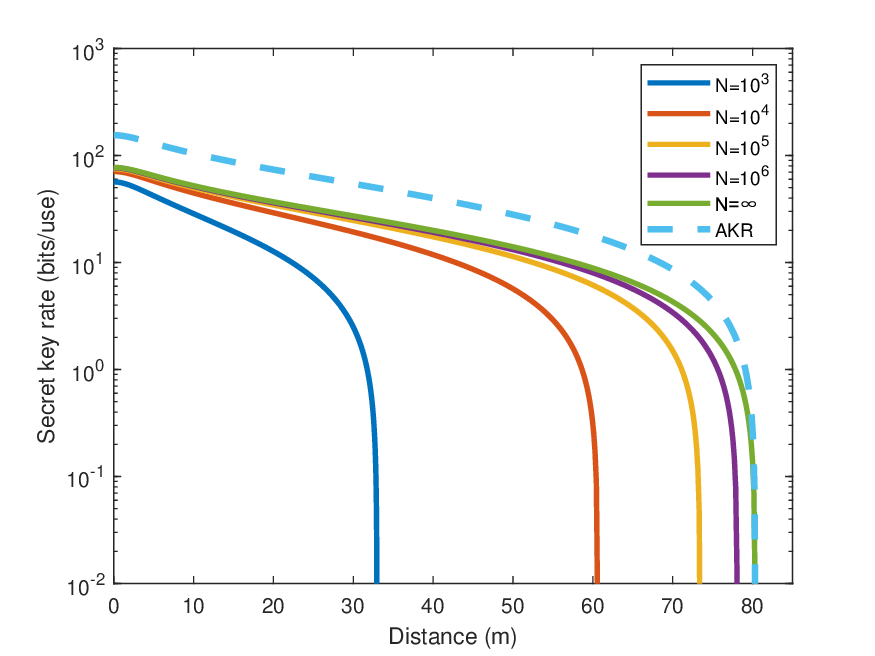}}
	\hfill	
	\subfloat[$N_{T_{A,B}} \times N_{R_{A,B}}=64\times64$]{\includegraphics[width=0.24\textwidth]{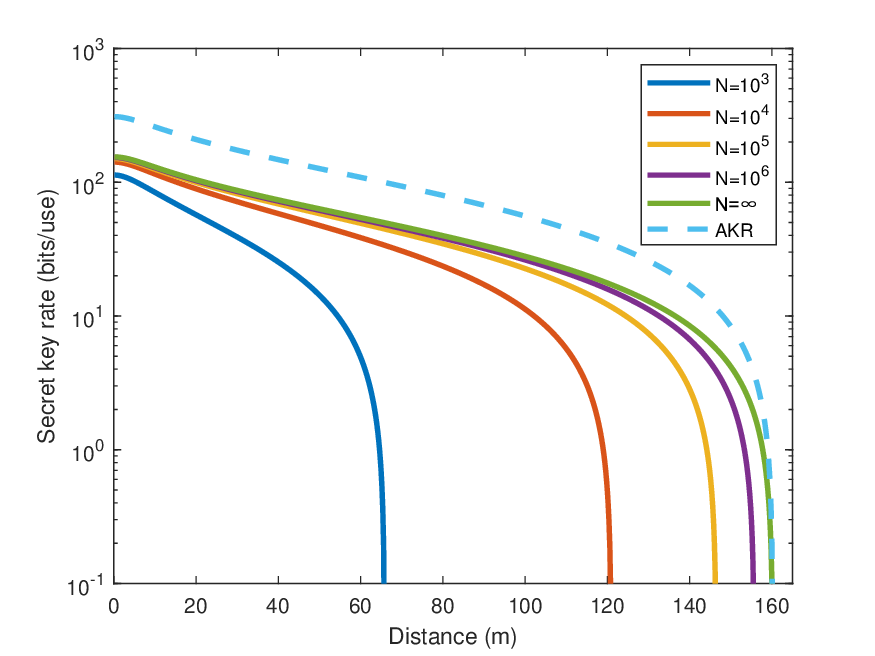}}
	\hfill 	
	\newline
	\subfloat[ $N_{T_{A,B}} \times N_{R_{A,B}}=128\times128$]{\includegraphics[width=0.24\textwidth]{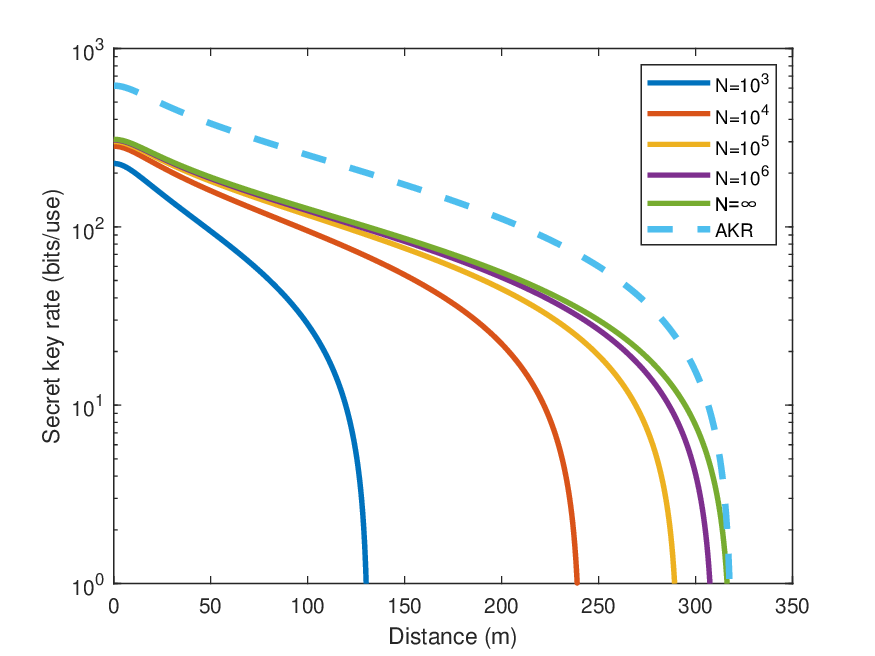}}
	\hfill 	
	\subfloat[$N_{T_{A,B}}\times N_{R_{A,B}}=256\times256$]{\includegraphics[width=0.24\textwidth]{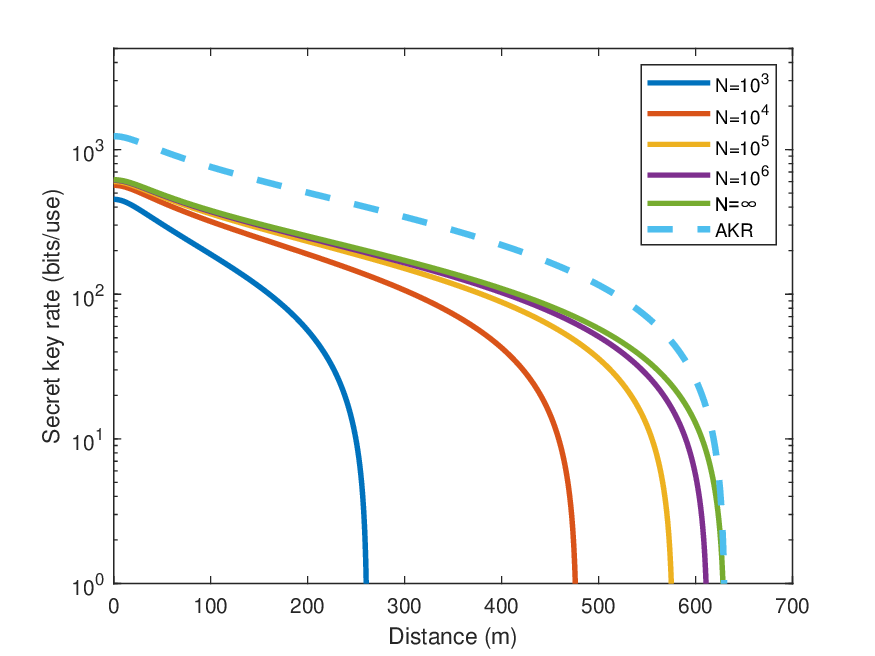}}
	\hfill 	
	\subfloat[$N_{T_{A,B}} \times N_{R_{A,B}}=512\times512$]{\includegraphics[width=0.24\textwidth]{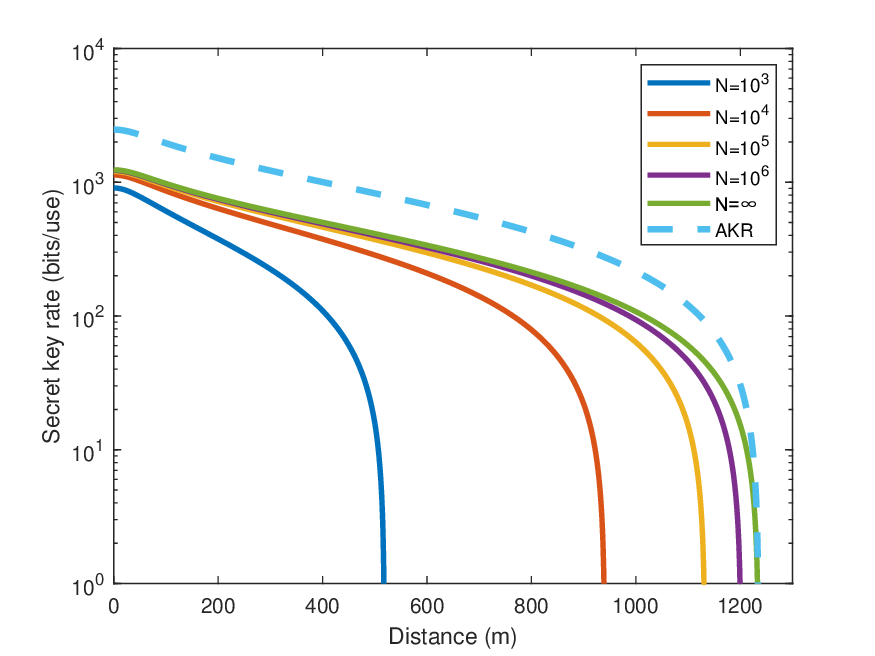}}
	\hfill	
	\subfloat[$N_{T_{A,B}} \times N_{R_{A,B}}=1024\times1024$]{\includegraphics[width=0.24\textwidth]{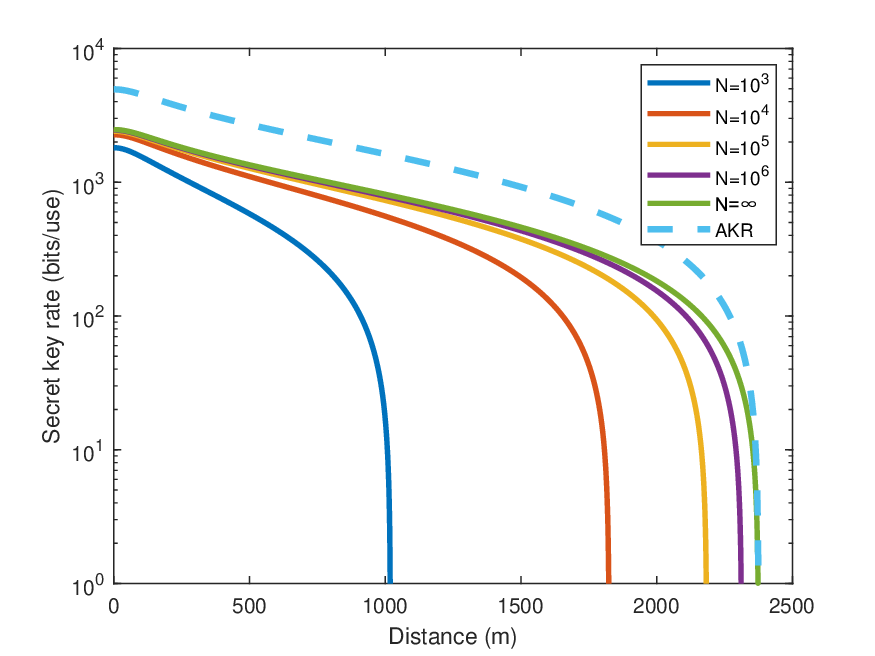}}
	\hfill 	
	\caption{Finite code rate as a function of transmission distance for the different MIMO configurations with different data block sizes. When the data block size is below $2\times 10^5$, the effective transmission distance of the CVMDI MIMO protocol decreases significantly. However, when the data block size reaches $2\times 10^6$, the effective transmission distance approaches that of an infinite range, offering an optimal balance between cost and performance. {The AKR curve represents the asymptotic key rate for this MIMO configuration.} $M=2N$, $f_c=100$GHz,  $\eta_{D_{A(B)_{i}}}=1$, {$\delta_{1,2}=0.6$dB/km}.}
	\label{fig6}
\end{figure*}
\begin{figure}[!t]
	\centering
	\includegraphics[width=0.8\linewidth]{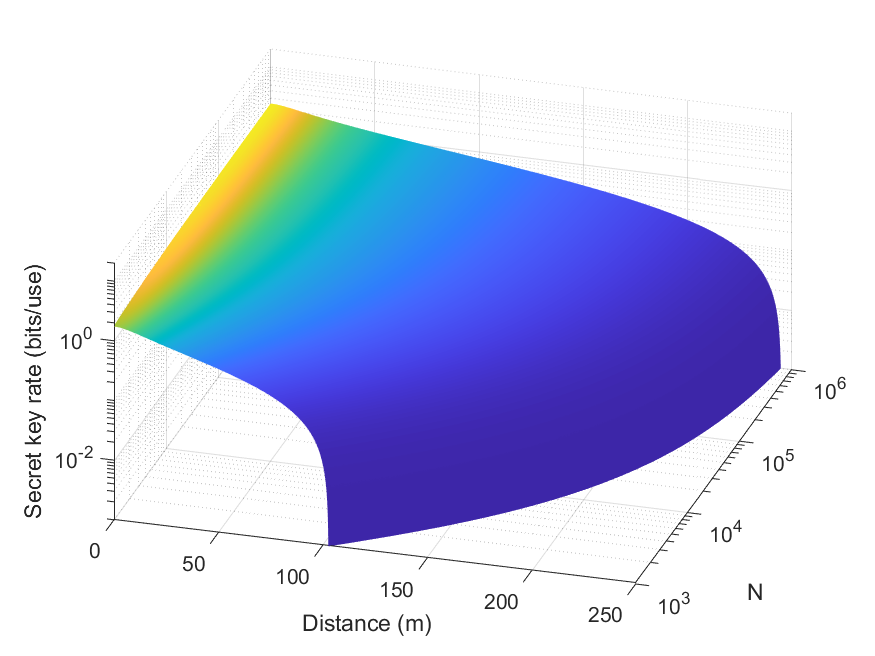}
	\caption{A three-dimensional surface plot illustrating the relationship between $K_{\mathrm{MIMO}}^{Fr}$, valid data block size $N$ and the transmission distance. $M=2N$, $f_c=100$GHz, $\eta_{D_{A(B)_{i}}}=1$, {$\delta_{1,2}=0.6$dB/km}.}
	\label{fig7}
\end{figure}

{Based on the derivation of Eq. (23),} we now have sufficient information to evaluate the impact of finite code on both MDI MIMO and SISO protocols.  In our analysis, we assume that the data block size used by Alice and Bob during the parameter estimation phase is half of the total data block size, i.e. $l=N=0.5M$. As determined in the previous subsection, the optimal frequency for long distance transmission in this protocol is 100 GHz, so we adopt $f_c$=100 GHz. We continue to restrict our analysis to the symmetric models. {All simulation parameters are outlined in Table I.} 

Figure \ref{fig6} shows the maximum transmission distance as a function of the finite code rate $K_{\mathrm{MIMO}}^{Fr}$ at 0.1 THz and room temperature, for various MIMO configurations and block sizes. The lower subplots display configurations of $8\times8$, $16\times16$, $32\times32$ and $64\times64$, while the upper subplots include $128\times128$, $256\times256$, $512\times512$ and $1024\times1024$.  In each subplot, a discernible trend emerges: as the block size increases, both the key rate and the maximum achievable transmission distance show a corresponding improvement. {This is because, as the block size increases, the estimators $T_{{LA(B)}_i}$ and $W_{{UA(B)}_i}$ in Eqs. (33) and (35) become closer to the true values $T_{{A(B)}_i}$ and $W_{{A(B)}_i}$. At the same time, the amount of data allocated for privacy amplification (see Eq. 24) decreases,  leading to an increase in the key rate of the proposed protocol. When the block size approaches infinity, Alice and Bob can accurately estimate the channel parameters and the data reserved for privacy amplification becomes negligible. However, since only a fraction $\frac{N}{M}$  of the all signals are actually encoded, the secret key rate under finite-size conditions remains lower than the asymptotic key rate. Notably,  when the block size reaches $M=2\times10^6$, the performance in terms of secret key rate and transmission distance  approaches that of an infinite block size. Therefore,  a block size of $M=2\times10^6$ provides a favorable balance between practical constraints and performance, making it well-suited for the implementation of the CVMDI MIMO protocol. }

For completeness, we also include a three-dimensional graph for SISO in Figure \ref{fig7}, which illustrates the relationship between block size, transmission distance and key rate.  As expected, the finite code effect reduces the maximum transmission distance of the SISO protocol and this limitation can be mitigated by increasing the block size. In comparison with Figure \ref{fig6}, it is evident that the SISO protocol consistently underperforms relative to the MIMO protocol under identical block size conditions.  This further emphasizes the importance of antenna technology in enhancing system performance.

\begin{table*}
	
	\centering
	{
		\renewcommand{\arraystretch}{1.5} 
		\caption{\textbf{Performance Comparison Between MDI MIMO and Existing MIMO-Based QKD Protocols}}
		\begin{tabular}{cccccc}
			\toprule
			Document & Transmission distance & MIMO configuration &Temperature &Frequency & Finite code \\
			\midrule
			\cite{33}  & 250m        & $1024\times1024$  & room temperature & 15THz  &  NO \\
			\cite{34}  & close 100m  & $256\times256$ & room temperature   & 15THz  & least-square \\
			\cite{35}  & over 100m   & $512\times512$  & room temperature  & 15THz &  least-square \\
			\cite{100} & 130m        & $512\times512$  & low temperature  & 15THz &  NO \\
			\cite{101} & over 100m   & $256\times256$  & room temperature  & 15THz  &  least-square \\
			\cite{99999} & over 500m   & $8\times8$  & room temperature  & 15THz  &  NO \\
			\cite{45} &  over 8000m  & $1024\times1024$ & low temperature & 0.1THz  &  NO\\
			\cite{45} &  over 3000m  & $1024\times1024$ & low temperature & 0.2THz  & NO\\
			MDI MIMO &  20m  & $8\times8$ & room temperature & 0.1THz  &  NO\\
			MDI MIMO &  2374m  & $1024\times1024$ & room temperature & 0.1THz  &  NO\\
			MDI MIMO &  125m  & $1024\times1024$ & room temperature & 1THz  &  NO\\
			MDI MIMO &  316m  & $128\times128 $& room temperature & 0.1THz  &  NO\\
			MDI MIMO &  307m  &$ 128\times128  $& room temperature& 0.1THz  &  MLE\\

			\bottomrule
		\end{tabular}
	} 
\end{table*}

\subsection{Comparison with other  literature}
To comprehensively assess the performance of the proposed CVMDI MIMO protocol, we compare it with several representative CVQKD schemes reported in recent literature, as summarized in Table~II. The comparison criteria include transmission distance, MIMO configuration, atmospheric temperature, operating frequency and the consideration of finite-size effects. Our protocol supports secure communication over long distances in room temperature. Specifically, under a 0.1THz frequency and a 1024$\times$1024 MIMO configuration, it enables secure key distribution over a distance of 2374m. Although prior work \cite{45} demonstrated long-distance transmission using MIMO-based CVQKD, its implementation requires a low-temperature environment, and the lack of an MDI framework leaves it vulnerable to detector-side attacks. It is noteworthy that the CVQKD protocol based on OTFS-MIMO achieves transmission over 500 meters even with a low-configuration MIMO setup, demonstrating its robustness against the severe path loss \cite{99999}. This result underscores the potential of combining advanced multi-carrier waveforms with spatial diversity to enable long-distance quantum communications in practical scenarios.

{A key distinguishing feature of our protocol is its MDI architecture, which inherently mitigates all detector-side channel attacks—an essential security guarantee absent in the compared literature. Additionally, our analysis incorporates finite-size effects using the MLE method, further enhancing the practicality and real-world applicability of the proposed scheme.}

 \section{CONCLUSION}
 
We have discussed the MIMO CVMDI QKD system operating at THz frequencies and provided a detailed description of the communication steps for both the EB and PM schemes. The secret key rate of the asymptotic and finite code for the proposed protocol is derived under the  Gaussian collective attacks. The protocol is designed to effectively minimize noise and other adverse factors, thereby ensuring consistent performance in complex environments.  The relevant analysis provides a solid theoretical foundation for understanding the protocol and ensures the reliability and security of the system.

The MIMO architecture enhances the MDI THz system's resilience to jamming and ensures reliable key transmission in free-space scenarios. {By adjusting key parameters, including the THz wave frequency and homodyne detection efficiency, the proposed protocol can be flexibly adapted to QKD systems with different channel lengths.} By appropriately configuring the number of antenna nodes in the CVMDI MIMO protocol, the proposed protocol can be effectively applied to both indoor and outdoor wireless communication. When the data block size is set to $2\times 10^6 $, nearly infinite block size performance can be achieved at a low cost,  which greatly enhances the practicality and economic feasibility of MIMO MDI THz QKD.


\end{document}